\documentclass[aps,prl,twocolumn,superscriptaddress,showpacs]{revtex4}

\usepackage{graphicx}
\usepackage{dcolumn}
\usepackage{bm}
\begin{document}

\title{Magnetization switching by ultrashort acoustic pulses}

\author{Oleksandr Kovalenko, Thomas Pezeril and Vasily V. Temnov$^{\ast}$}
\affiliation{Institut des Mol\'ecules et Mat\'eriaux du Mans, UMR
CNRS 6283, Universit\'e du Maine, 72085 Le Mans cedex, France}

\date{\today}

\email{vasily.temnov@univ-lemans.fr}


\begin{abstract}
It is shown theoretically that a single a few picoseconds long
acoustic pulse can reverse magnetization in a magneto-strictive
material Terfenol-D. Following giant magneto-elastic changes of
free energy density the magnetization vector is kicked out of a
local in-plane energy mininum and decays into another minimum. For
acoustic pulse duration significantly shorter than magnetization
precession period $\tau_{ac}\ll T_{prec}$, the switching threshold
is determined by the {\it acoustic pulse area}, i.e. pulse
integral in time domain, similar to coherent phenomena in optics.
Simulation results are summarized in a magneto-acoustic switching
diagram and discussed in the context of all-optical magnetization
switching by circularly polarized light pulses.
\end{abstract}


\maketitle

Searching for new possibilities of ultrafast magnetization
switching is motivated by continuously growing demand in faster
data recording technologies, which are based on reversal of
individual magnetic bits at the nano-scale. Among different
mechanisms of magnetization switching the acoustically driven
switching at ultrafast time scales remains largely unexplored.

The first pioneering time-resolved observation of magnetization
dynamics in ferromagnetic nickel induced by femtosecond laser
pulses revealed the phenomenon of ultrafast demagnetization in
nickel \cite{Beaurepaire96PRL76}. The ultrafast drop of
magnetization on a sub-picosecond time scale was caused by
transient elevation of electron temperature close to the Curie
point. The subsequent dynamics of magnetization recovery on a time
scale exceeding tens of picoseconds could be adequately reproduced
by Landau-Lifshitz-Gilbert (LLG) equations describing damped
precession of magnetization vector in the presence of
temperature-dependent magneto-crystalline anisotropy
\cite{Bigot05CP318}.

The most recent experiments combining picosecond acoustics
\cite{Thomsen86PRB34} with ultrafast magneto-optics showed that
the magneto-crystalline anisotropy can be also changed by
picosecond acoustic strain pulses to a ferromagnetic semiconductor
GaMnAs \cite{Scherbakov10PRL105,Thevenard10PRB82} or nickel
\cite{Kim12PRL109}, thus triggering the magnetization precession
without heating the sample. However, due to the relatively small
magneto-strictive coefficient in GaMnAs ($\Lambda\sim 10^{-5}$ at
cryogenic temperature \cite{Masmanidis05PRL95}) and nickel
($\Lambda\sim 3\times 10^{-5}$ at room temperature
\cite{ClarkHandbookPhysChem}), the resulting transient change in
magnetization direction appeared to be too small and the
magnetization vector returned to its initial direction, i.e.
magnetization switching (reversal) did not occur.

The first experimental demonstration of ultrafast {\it
non-thermal} magnetization switching was reported in 1998 by Back
and co-workers \cite{Back98PRL81}, who used ultrashort pulses of
magnetic field induced by relativistic electron bunches to switch
magnetization in Co/Pt film. Most recently a spectacular
observation of all-optical magnetization switching in GdFeCo using
circularly polarized light pulses
\cite{Stanciu07PRL99,Vahaplar09PRL103}, raised a lot of questions
about physically unclear switching mechanisms suggesting that not
only transient overheating of electrons but also the dynamics of
lattice temperature may be necessary to explain the underlying
physics \cite{Steil2011PRB84}. Not only the minimum amount of
deposited heat \cite{Steil2011PRB84} but also a minimum amount of
'circularity' was necessary to obtain switching
\cite{Alebrand2012PRB85}, at least within a certain range of
optical excitation \cite{Ostler2012Ncomm3}.

In this letter we theoretically investigate the interaction of
ultrashort acoustic pulses with Terfenol-D
(Tb$_{x}$Dy$_{1-x}$)Fe$_2$, the rare-earth compound famous through
its giant magneto-strictive coefficient $\Lambda\sim 10^{-3}$
\cite{ClarkHandbookPhysChem}, and demonstrate the possibility of
ultrafast magneto-acoustic switching. Moreover, the results are
discussed within the framework of recently observed all-optical
magnetization switching by single circularly polarized femtosecond
laser pulses, which is undoubtedly accompanied by the generation
of picosecond pulses of coherent acoustic phonons
\cite{Thomsen86PRB34}.

The phenomenological expression for free energy density
$F(\overrightarrow{M})=F_k+F_{me}+F_d+F_z$ for (110) thin films of
Tb$_{0.27}$Dy$_{0.73}$Fe$_2$ epitaxially grown on sapphire reads
\cite{delaFuente04JPCM16}:
\begin{eqnarray}
\label{mk-term}
F_k&=&K_1(\alpha^2_x\alpha^2_y+\alpha^2_x\alpha^2_z+\alpha^2_z\alpha^2_y)+K_2(\alpha^2_x\alpha^2_y\alpha^2_z)\\
\nonumber
\label{me-term}
F_{me}&=&b_1(\alpha^2_xe_{xx}+\alpha^2_ye_{yy}+\alpha^2_ze_{zz})+\\
&+&b_2(\alpha_x\alpha_ye_{xy}+ \alpha_x\alpha_ze_{xz}+
\alpha_y\alpha_ze_{yz})\,\\
\label{ms-term} F_d&=&\frac{\mu_0}{2}(M_s\cos\theta)^2\,,
\end{eqnarray}
where $\alpha_x,\alpha_y,\alpha_z$ are direction cosines of the
magnetization vector
$\overrightarrow{M}=M_s(\alpha_x,\alpha_y,\alpha_z)$ in the
crystallographic coordinate system $(x,y,z)$ and saturation
magnetization $\mu_0M_s=0.945$~T. In the rotated frame
$(x^{\prime},y^{\prime},z^{\prime})$ the direction of
magnetization is determined by two angles $\theta$ (out-of-plane
angle) and $\phi$ (in-plane angle), see Fig.~1(a). In
Eq.~(\ref{mk-term}-\ref{ms-term}) $F_k$, $F_{me}$ and $F_d$ denote
the magneto-crystalline anisotropy, magneto-elastic and
magneto-static terms, respectively, and Zeeman contribution
$F_z=-\mu_0\overrightarrow{H}_{ext}\overrightarrow{M}$ is
disregarded through out this paper as we consider the case of zero
external magnetic field, $\overrightarrow{H}_{ext}=0$.

Epitaxial growth of a thin Terfenol-D film in (110) direction on a
lattice-mismatched substrate induces the built-in static strain
described by the following tensor:
\begin{equation}
\label{StaticStrainTensor} e_{stat}=\left( \begin{array}{ccc}
0 & e_{xy} & 0 \\
e_{xy} & 0 & 0 \\
0 & 0 & -\frac{2c_{12}}{c_{11}}e_{xy} \end{array} \right)\ \,,
\end{equation}
which is determined by a single strain component
$e_{xy}$=-0.55$\%$ \cite{Mougin00PRB62} in a crystallographic
coordinate frame $(x,y,z)$ (Fig.~1(a)). The competition of
different contributions in the total free energy density results
into four local in-plane energy minima corresponding to four
different magnetization directions 1,2,3,4 in Fig.~1(b). The
explicit dependence of the magneto-elastic term $F_{me}$ both on
the strain components and magnetization direction and large values
of magneto-elastic coupling coefficients
$b_1=-\frac{3}{2}\Lambda_{100}(c_{11}-c_{12})=-80$~MPa and
$b_2=-3\Lambda_{111}c_{44}=-85$~MPa \cite{FreeEnergyParameters}
suggest that application of external strain will shift the minima
of free energy minima and therefore, change the magnetization
direction. Indeed, the application of time-independent uniaxial
strain $\eta$ in the direction normal to the surface of a thin
Terfenol-D film results into the in-plane shift of all four energy
minima by the angle $\Delta\Phi$, as illustrated in Fig.~1(c) for
$\eta=-0.3\%$ (film compression) and $\eta=0.9\%$ (film tension).

The action of a time-dependent uniaxial strain (acoustic pulse)
$\eta(t)$ can be described by adding the following dynamic tensor
in a crystallographic coordinate frame
\begin{equation}
\label{DynamicStrainTensor} e_{dyn}(t)=\frac{1}{2}\left(
\begin{array}{ccc}
\eta(t) & -\eta(t) & 0 \\
-\eta(t) & \eta(t) & 0 \\
0 & 0 & 0  \end{array} \right)\ \,,
\end{equation}
where the rotation of coordinates system by 45$^{\circ}$ from
$(x^{\prime},y^{\prime},z^{\prime})$ into $(x,y,z)$ leads to
factor 1/2 and generates the non-diagonal terms.

Inserting the total strain $e(t)=e_{stat}+e_{dyn}(t)$ in
Eq.~(\ref{me-term}) generates explicit time-dependence of free
energy $F(t)$, which drives the magneto-acoustic dynamics
described by LLG equation \cite{Kim12PRL109}
\begin{equation}\label{LLGEquation}
\frac{d\overrightarrow{M}}{dt}=-\gamma\mu_0
(\overrightarrow{M}\times\overrightarrow{H}_{eff})
+\frac{\alpha}{M_s} \bigg (\overrightarrow{M}\times
\frac{d\overrightarrow{M}}{dt}\bigg )
\end{equation}
where the first term describes the torque driving the precession
of magnetization vector around the effective time-dependent
magnetic field $H_{eff}(t)$
\begin{equation}
\label{Heff}
\overrightarrow{H}_{eff}(t)=-\frac{1}{\mu_0}\frac{dF(t)}{d\overrightarrow{M}}
\end{equation}
and the second term describes precession damping according to the
phenomenological Gilbert damping parameter $\alpha=0.1$
\cite{Fashami11N22}; $\gamma$ is the gyromagnetic ratio.

\begin{figure}
\includegraphics[width=8 cm]{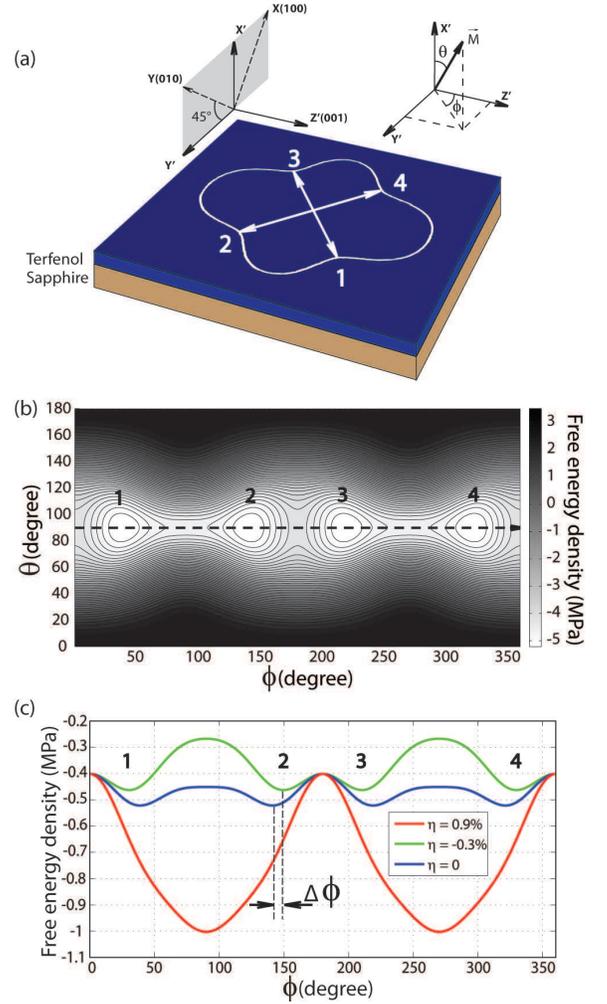}
\caption{Magneto-elastic free energy density in a thin (110) film
of Terfenol-D grown on sapphire (a) possesses four in-plane energy
minima (b). Application of a {\it static} uniaxial compressive
($\eta=-0.3\%$) or tensile ($\eta=0.9\%$) strain in the direction
perpendicular to the film leads to the in-plane shift $\Delta\Phi$
of all four energy minima (c).} \label{figure1}
\end{figure}

\begin{figure}
\includegraphics[width=8 cm]{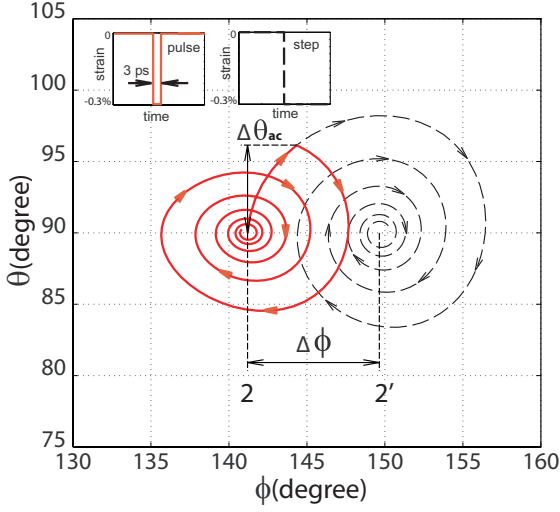}
\caption{Starting from energy minimum 2, a {\it dynamic} step-like
strain induces a damped magnetization precession around the
shifted minimum $2^{\prime}$ (dashed line). An ultrashort acoustic
pulse results into the out-of-plain kick $\Delta\theta_{ac}$
followed by damped precession into the initial minimum 2
(continuous line).} \label{fig:fig 2}
\end{figure}

If the system is initially prepared in minimum 2, the application
of instantaneous step-like strain \cite{AcPulseDuration} results
into damped precession of magnetization vector into the new
minimum $2^{\prime}$ with a precession period $T_{prec}\simeq
25$~ps, see Fig.~2. However, if the strain is turned off after
some time, the situation equivalent to the application of a
rectangular acoustic pulse of finite duration $\tau_{ac}$, the
precession trajectory will decay back into minimum 2. Such
magneto-acoustic precession trajectory induced by the action of a
picosecond acoustic pulse with $\tau_{ac}=3$~ps and strain
amplitude $\eta_{ac}=3\times 10^{-3}$ is shown in Fig.~2 and can
be explained analytically.

LLG equations (\ref{LLGEquation}) for the initial condition in one
of the four energy minima read
\begin{equation}
\label{Kick} \frac{d\phi}{dt}=0\,,\,
\frac{d\theta}{dt}=\gamma\frac{(b_2+2b_1)}{2M_s}\alpha^{\prime}_y\alpha^{\prime}_z\eta(t)
\end{equation}
and their integration for an ultrashort acoustic pulse $\eta(t)$
obeying the condition $\tau_{ac}\ll T_{prec}$ approximate well the
acoustic {\it out-of-plane kick} of the magnetization vector by
the angle
\begin{equation}
\label{Kick}
\Delta\theta_{ac}\simeq\gamma\frac{(b_2+2b_1)}{2M_s}\alpha^{\prime}_y\alpha^{\prime}_z\int\eta(t)dt\,.
\end{equation}
The product of directional cosines
$\alpha^{\prime}_y\alpha^{\prime}_z$ in Eq.~({\ref{Kick}) equals
0.48 for energy minima 1 and 3 and -0.48 for two other minima,
respectively. Therefore, depending on the initial condition, the
same acoustic pulse will kick the magnetization vector out of
sample plane in opposite directions. Equation (\ref{Kick})
clarifies the microscopic physical model beyond the time-dependent
magnetic torque $|\overrightarrow{M}\times
\overrightarrow{H_{eff}}|$ introduced by Kim and co-workers
\cite{Kim12PRL109} and shows that the prefactor in
Eq.~(\ref{Kick}) is dominated by the ratio of magneto-elastic
coupling coefficients $b_1$ and $b_2$ (which are both proportional
to the respective coefficients of magneto-strictive tensor
$\Lambda$) to saturation magnetization $M_s$.

Similar to polarization dynamics in coherent optics, the acoustic
rotation angle $\Delta\theta_{ac}$ of magnetization vector appears
to be proportional to the {\it acoustic pulse area}
$\int{\eta(t)dt}$ for arbitrary acoustic pulses obeying
$\tau_{ac}\ll T_{prec}\simeq 25$~ps. It suggests that the
so-called bi-polar acoustic pulses generated at free metal-air
interfaces \cite{Thomsen86PRB34} are particularly inefficient in
magneto-acoustics since positive and negative parts in a bi-polar
pulse cancel each other giving zero acoustic pulse area. Recently
observed large-amplitude unipolar acoustic pulses with 3~ps
duration and amplitudes up to $1\%$ generated in a thin cobalt
transducer sandwiched between dielectric substrate and a layer of
noble metal \cite{Temnov12NPhoton6,TemnovAcoustoPlasmonics} are
better suited for experimental investigations in coherent
magneto-acoustics.

When using rectangular unipolar acoustic pulses the kick angle
$\Delta\theta_{ac}$ is proportional to the product
$\eta_{ac}\times\tau_{ac}$ and thus can be increased by using
larger strain amplitude $\eta_{ac}$ or somewhat longer pulse
duration $\tau_{ac}\ll T_{prec}$.

\begin{figure}
\includegraphics[width=8 cm]{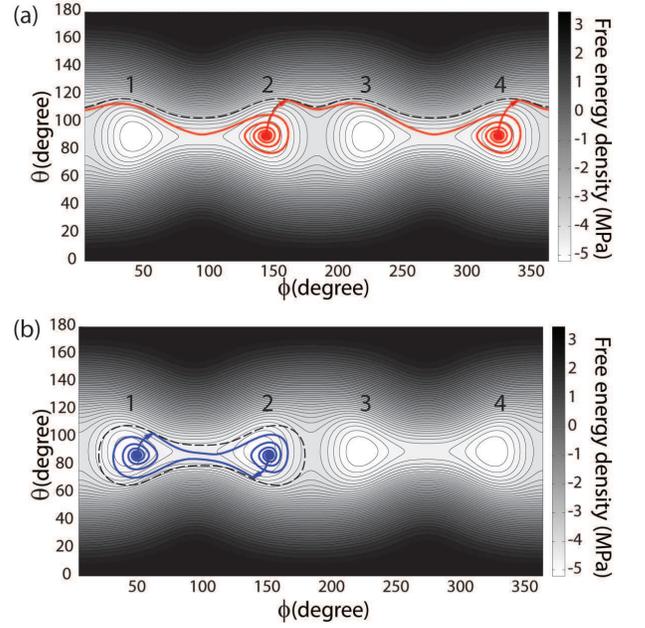}
\caption{After application of large-amplitude 3~ps long acoustic
pulse the magnetization vector initially prepared in state 2 will
decay (switch) into one of the neighboring energy minima. (a) a
sequence of compressional pulses with $\eta_{ac}=-1.6\%$, induces
the clock-wise rotation of magnetization vector: $2\rightarrow 4$,
$4\rightarrow 2$, etc. (b) a sequence of tensile pulses with
$\eta_{ac}=0.9\%$ results into repetitive switching between two
adjacent energy minima: $2\rightarrow 1$, $1\rightarrow 2$, etc.
Dashed contours show the magnetization trajectories after
excitation by a single pulse in the absence of damping.}
\label{fig:fig 3}
\end{figure}

Figure 3 shows that both compressional ($\eta_{ac}<0$, Fig.~3a) or
tensile ($\eta_{ac}>0$, Fig.~3b) rectangular unipolar acoustic
pulses with $\tau_{ac}=3$~ps are capable of switching the initial
magnetization direction into the new minimum, which represents the
main result of this paper. For example, a three picosecond long
compressional pulse with strain amplitude 1.6$\%$ switches the
magnetization from minimum 2 into minimum 4 ($2\rightarrow 4$, see
Fig.~3a). Application of a consequent identical compressional
pulse bring the magnetization back into minimum 2 ($4\rightarrow
2$, suggesting that a pulse train will result into clock-wise
rotation of magnetization vector. In contrast, a sequence of 3~ps
long tensile strain pulses will periodically switch the
magnetization between minima 2 and 1 ($2\rightarrow 1$ and
$1\rightarrow 2$ see Fig.~3b).

Therefore the results of our simulations suggest that a clean
experimental demonstration of magneto-acoustic switching would
necessarily imply a single-shot experiment where the magnetic
system is prepared in the same state before the action of the
consequent acoustic pulse.

The typical kick angle required for switching is about
20$^{\circ}$ and the different threshold switching amplitudes and
pathways for tensile and compressive pulses are caused by
different hight of the potential barrier between the neighboring
energy minima. The more general phase diagram for magneto-acoustic
switching is shown in Fig.~4, where the boundaries between
different switching zones generally follow the $1/\tau_{ac}$
dependence, in agreement with the assumption that primarily the
amplitude of out-of-plane acoustic kick
$\Delta\theta_{ac}\sim\eta_{ac}\times\tau_{ac}=const$ determines
the switching pathway. Similar analysis for the acoustic shear
pulses leads to the same conclusions, in particular with respect
to the acoustic pulse area and dependence of the switching
amplitude on the acoustic pulse duration.

\begin{figure}
\includegraphics[width=8 cm]{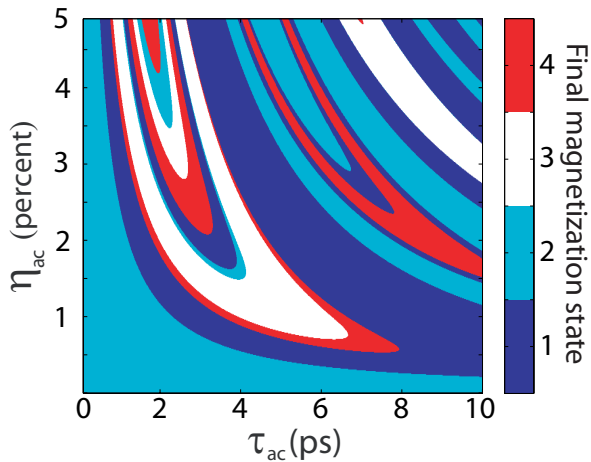}
\caption{Diagram of magneto-acoustic switching. Depending on the
duration $\tau_{ac}$ and amplitude $\eta_{ac}$ of a unipolar
tensile acoustic pulse, the magnetization vector initially
prepared in state 2 undergoes the transition $2\rightarrow 1$
(dark blue area), $2\rightarrow 4$ (red area), $2\rightarrow 3$
(white area) or $2\rightarrow 2$ (no switching, light blue area).}
\label{fig:fig 4}
\end{figure}

It is quite remarkable that the threshold fluence for all-optical
magnetization switching in a rare-earth compound GdFeCo induced by
a single circularly polarized pump pulse almost does not depend on
the optical pulse duration \cite{Steil2011PRB84}. Longer optical
excitation leads to longer acoustic pulses \cite{Dehoux06JAP100},
which are generated by thermo-elastic expansion of laser-heated
GdFeCo. The analysis of thermo-elastic generation mechanism by
longer optical pulses suggests that the absorbed laser fluence is
proportional to the product of acoustic pulse amplitude by its
duration, i.e to the acoustic pulse area. Therefore, the observed
constant threshold fluence for all-optical magnetization switching
in GdFeCo could be explained by magneto-elastic mechanism. From
the magneto-acoustic point of view a significantly lower
magnetostriction coefficient $\Lambda\sim 10^{-5}$ in GdFeCo
\cite{Yoshino88JAP64} is balanced by a much lower room temperature
saturation magnetization $\mu_0M_s\simeq 8$~mT close to the
compensation point \cite{Stanciu07PRL99}, thus giving the same
order-of-magnitude prefactor $\Lambda/M_s$ in Eq.~(\ref{Kick}).
The observed increase of switching fluence in GdFeCo with
temperature \cite{Alebrand2012PRB85} is consistent with the
decrease of $\Lambda/M_s$ in rare-earth compounds for higher
temperatures \cite{delaFuente04JPCM16}. Therefore, despite of
being far away from quantitative modeling, all these arguments
corroborate the hypothesis \cite{Temnov12NPhoton6} that ultrashort
acoustic phonon pulses may play an important role in the dynamics
of all-optical magnetization switching in GdFeCo
\cite{Stanciu07PRL99,Vahaplar09PRL103}. Moreover, possible
magneto-elastic contributions of helically polarized acoustic
shear pulses excited by circularly polarized electromagnetic
pulses \cite{Dobbs69JPCS31,Thomas68PRL20} may solve the puzzle of
an undefined long-living reservoir for angular momentum as
highlighted in the most recent systematic investigations
\cite{Steil2011PRB84,Alebrand2012PRB85}.

To summarize, in this letter we have theoretically predicted the
new mechanism of ultrafast {\it non-thermal} magneto-acoustic
switching in Terfenol-D. This phenomenon may open the door to
ultrafast magnetic recording not relying on heating the magnetic
material close the Curie point.

Stimulating discussions with Stephane Andrieu and Karine Dumesnil
and the financial support by {\it Nouvelle \'{e}quipe, nouvelle
th\'{e}matique de la R\'{e}gion Pays de La Loire} are gratefully
acknowledged.



\begin{thebibliography}{23}
\expandafter\ifx\csname
natexlab\endcsname\relax\def\natexlab#1{#1}\fi
\expandafter\ifx\csname bibnamefont\endcsname\relax
  \def\bibnamefont#1{#1}\fi
\expandafter\ifx\csname bibfnamefont\endcsname\relax
  \def\bibfnamefont#1{#1}\fi
\expandafter\ifx\csname citenamefont\endcsname\relax
  \def\citenamefont#1{#1}\fi
\expandafter\ifx\csname url\endcsname\relax
  \def\url#1{\texttt{#1}}\fi
\expandafter\ifx\csname
urlprefix\endcsname\relax\def\urlprefix{URL }\fi
\providecommand{\bibinfo}[2]{#2}
\providecommand{\eprint}[2][]{\url{#2}}

\bibitem[{\citenamefont{Beaurepaire et~al.}(1996)\citenamefont{Beaurepaire,
  Merle, Daunois, and Bigot}}]{Beaurepaire96PRL76}
\bibinfo{author}{\bibfnamefont{E.}~\bibnamefont{Beaurepaire}},
  \bibinfo{author}{\bibfnamefont{J.~C.} \bibnamefont{Merle}},
  \bibinfo{author}{\bibfnamefont{A.}~\bibnamefont{Daunois}}, \bibnamefont{and}
  \bibinfo{author}{\bibfnamefont{J.-Y.} \bibnamefont{Bigot}},
  \bibinfo{journal}{Phys.\ Rev.\ Lett.} \textbf{\bibinfo{volume}{76}},
  \bibinfo{pages}{4250} (\bibinfo{year}{1996}).

\bibitem[{\citenamefont{Bigot et~al.}(2005)\citenamefont{Bigot, Vomir, Andrade,
  and Beaurepaire}}]{Bigot05CP318}
\bibinfo{author}{\bibfnamefont{J.-Y.} \bibnamefont{Bigot}},
  \bibinfo{author}{\bibfnamefont{M.}~\bibnamefont{Vomir}},
  \bibinfo{author}{\bibfnamefont{L.~J.~F.} \bibnamefont{Andrade}},
  \bibnamefont{and}
  \bibinfo{author}{\bibfnamefont{E.}~\bibnamefont{Beaurepaire}},
  \bibinfo{journal}{Chemical Physics} \textbf{\bibinfo{volume}{318}},
  \bibinfo{pages}{137} (\bibinfo{year}{2005}).

\bibitem[{\citenamefont{Thomsen et~al.}(1986)\citenamefont{Thomsen, Grahn,
  Maris, and Tauc}}]{Thomsen86PRB34}
\bibinfo{author}{\bibfnamefont{C.}~\bibnamefont{Thomsen}},
  \bibinfo{author}{\bibfnamefont{H.~T.} \bibnamefont{Grahn}},
  \bibinfo{author}{\bibfnamefont{H.~J.} \bibnamefont{Maris}}, \bibnamefont{and}
  \bibinfo{author}{\bibfnamefont{J.}~\bibnamefont{Tauc}},
  \bibinfo{journal}{Phys.\ Rev.\ B} \textbf{\bibinfo{volume}{34}},
  \bibinfo{pages}{4129} (\bibinfo{year}{1986}).

\bibitem[{\citenamefont{Scherbakov et~al.}(2010)\citenamefont{Scherbakov,
  Salasyuk, Akimov, Liu, Bombeck, Brueggeman, Yakovlev, Sapega, Furdyna, and
  Bayer}}]{Scherbakov10PRL105}
\bibinfo{author}{\bibfnamefont{A.~V.} \bibnamefont{Scherbakov}},
  \bibinfo{author}{\bibfnamefont{A.~S.} \bibnamefont{Salasyuk}},
  \bibinfo{author}{\bibfnamefont{A.~V.} \bibnamefont{Akimov}},
  \bibinfo{author}{\bibfnamefont{X.}~\bibnamefont{Liu}},
  \bibinfo{author}{\bibfnamefont{M.}~\bibnamefont{Bombeck}},
  \bibinfo{author}{\bibfnamefont{C.}~\bibnamefont{Brueggeman}},
  \bibinfo{author}{\bibfnamefont{D.~R.} \bibnamefont{Yakovlev}},
  \bibinfo{author}{\bibfnamefont{V.~F.} \bibnamefont{Sapega}},
  \bibinfo{author}{\bibfnamefont{J.~K.} \bibnamefont{Furdyna}},
  \bibnamefont{and} \bibinfo{author}{\bibfnamefont{M.~.} \bibnamefont{Bayer}},
  \bibinfo{journal}{Phys.\ Rev.\ Lett.} \textbf{\bibinfo{volume}{105}},
  \bibinfo{pages}{117204} (\bibinfo{year}{2010}).

\bibitem[{\citenamefont{Thevenard et~al.}(2010)\citenamefont{Thevenard,
  Perrone, Gourdon, Testelin, Cubukcu, Charron, Vincent, Lemaitre, and
  Perrin}}]{Thevenard10PRB82}
\bibinfo{author}{\bibfnamefont{L.}~\bibnamefont{Thevenard}},
  \bibinfo{author}{\bibfnamefont{E.}~\bibnamefont{Perrone}},
  \bibinfo{author}{\bibfnamefont{C.}~\bibnamefont{Gourdon}},
  \bibinfo{author}{\bibfnamefont{C.}~\bibnamefont{Testelin}},
  \bibinfo{author}{\bibfnamefont{M.}~\bibnamefont{Cubukcu}},
  \bibinfo{author}{\bibfnamefont{E.}~\bibnamefont{Charron}},
  \bibinfo{author}{\bibfnamefont{S.}~\bibnamefont{Vincent}},
  \bibinfo{author}{\bibfnamefont{A.}~\bibnamefont{Lemaitre}}, \bibnamefont{and}
  \bibinfo{author}{\bibfnamefont{B.}~\bibnamefont{Perrin}},
  \bibinfo{journal}{Phys.\ Rev.\ B} \textbf{\bibinfo{volume}{82}},
  \bibinfo{pages}{104422} (\bibinfo{year}{2010}).

\bibitem[{\citenamefont{Kim et~al.}(2012)\citenamefont{Kim, Vomir, and
  Bigot}}]{Kim12PRL109}
\bibinfo{author}{\bibfnamefont{J.~W.} \bibnamefont{Kim}},
  \bibinfo{author}{\bibfnamefont{M.}~\bibnamefont{Vomir}}, \bibnamefont{and}
  \bibinfo{author}{\bibfnamefont{J.-Y.} \bibnamefont{Bigot}},
  \bibinfo{journal}{Phys.\ Rev.\ Lett.} \textbf{\bibinfo{volume}{109}},
  \bibinfo{pages}{166601} (\bibinfo{year}{2012}).

\bibitem[{\citenamefont{Masmanidis et~al.}(2005)\citenamefont{Masmanidis, Tang,
  Myers, Li, Greve, Vermeulen, Roy, and Roukes}}]{Masmanidis05PRL95}
\bibinfo{author}{\bibfnamefont{S.~C.} \bibnamefont{Masmanidis}},
  \bibinfo{author}{\bibfnamefont{H.~X.} \bibnamefont{Tang}},
  \bibinfo{author}{\bibfnamefont{E.~B.} \bibnamefont{Myers}},
  \bibinfo{author}{\bibfnamefont{M.}~\bibnamefont{Li}},
  \bibinfo{author}{\bibfnamefont{K.~D.} \bibnamefont{Greve}},
  \bibinfo{author}{\bibfnamefont{G.}~\bibnamefont{Vermeulen}},
  \bibinfo{author}{\bibfnamefont{W.~V.} \bibnamefont{Roy}}, \bibnamefont{and}
  \bibinfo{author}{\bibfnamefont{M.~L.} \bibnamefont{Roukes}},
  \bibinfo{journal}{Phys.\ Rev.\ Lett.} \textbf{\bibinfo{volume}{95}},
  \bibinfo{pages}{187306} (\bibinfo{year}{2005}).

\bibitem[{\citenamefont{Clark}(1982)}]{ClarkHandbookPhysChem}
\bibinfo{author}{\bibfnamefont{A.~E.} \bibnamefont{Clark}}, in
  \emph{\bibinfo{booktitle}{Handbook of the Physics and Chemistry of Rare
  Earth}}, edited by \bibinfo{editor}{\bibfnamefont{K.~A.}
  \bibnamefont{Gschneider}} \bibnamefont{and}
  \bibinfo{editor}{\bibfnamefont{L.}~\bibnamefont{Eyring}}
  (\bibinfo{publisher}{Amsterdam: North Holland}, \bibinfo{year}{1982}).

\bibitem[{\citenamefont{Back et~al.}(1998)\citenamefont{Back, Weller, Heidmann,
  Mauri, Guarisco, Garwin, and Siegmann}}]{Back98PRL81}
\bibinfo{author}{\bibfnamefont{C.~H.} \bibnamefont{Back}},
  \bibinfo{author}{\bibfnamefont{D.}~\bibnamefont{Weller}},
  \bibinfo{author}{\bibfnamefont{J.}~\bibnamefont{Heidmann}},
  \bibinfo{author}{\bibfnamefont{D.}~\bibnamefont{Mauri}},
  \bibinfo{author}{\bibfnamefont{D.}~\bibnamefont{Guarisco}},
  \bibinfo{author}{\bibfnamefont{E.~L.} \bibnamefont{Garwin}},
  \bibnamefont{and} \bibinfo{author}{\bibfnamefont{H.~C.}
  \bibnamefont{Siegmann}}, \bibinfo{journal}{Phys.\ Rev.\ Lett.}
  \textbf{\bibinfo{volume}{81}}, \bibinfo{pages}{3251} (\bibinfo{year}{1998}).

\bibitem[{\citenamefont{Stanciu et~al.}(2007)\citenamefont{Stanciu, Hansteen,
  Kimel, Kirilyuk, Tsukamoto, Itoh, and Rasing}}]{Stanciu07PRL99}
\bibinfo{author}{\bibfnamefont{C.~D.} \bibnamefont{Stanciu}},
  \bibinfo{author}{\bibfnamefont{F.}~\bibnamefont{Hansteen}},
  \bibinfo{author}{\bibfnamefont{A.~V.} \bibnamefont{Kimel}},
  \bibinfo{author}{\bibfnamefont{A.}~\bibnamefont{Kirilyuk}},
  \bibinfo{author}{\bibfnamefont{A.}~\bibnamefont{Tsukamoto}},
  \bibinfo{author}{\bibfnamefont{A.}~\bibnamefont{Itoh}}, \bibnamefont{and}
  \bibinfo{author}{\bibfnamefont{T.}~\bibnamefont{Rasing}},
  \bibinfo{journal}{Phys.\ Rev.\ Lett.} \textbf{\bibinfo{volume}{99}},
  \bibinfo{pages}{047601} (\bibinfo{year}{2007}).

\bibitem[{\citenamefont{Vahaplar et~al.}(2009)\citenamefont{Vahaplar,
  Kalashnikova, Kimel, Hinzke, Nowak, Chantrell, Tskamoto, Itoh, Kirilyuk, and
  Rasing}}]{Vahaplar09PRL103}
\bibinfo{author}{\bibfnamefont{K.}~\bibnamefont{Vahaplar}},
  \bibinfo{author}{\bibfnamefont{A.~M.} \bibnamefont{Kalashnikova}},
  \bibinfo{author}{\bibfnamefont{A.~V.} \bibnamefont{Kimel}},
  \bibinfo{author}{\bibfnamefont{D.}~\bibnamefont{Hinzke}},
  \bibinfo{author}{\bibfnamefont{U.}~\bibnamefont{Nowak}},
  \bibinfo{author}{\bibfnamefont{R.}~\bibnamefont{Chantrell}},
  \bibinfo{author}{\bibfnamefont{A.}~\bibnamefont{Tskamoto}},
  \bibinfo{author}{\bibfnamefont{A.}~\bibnamefont{Itoh}},
  \bibinfo{author}{\bibfnamefont{A.}~\bibnamefont{Kirilyuk}}, \bibnamefont{and}
  \bibinfo{author}{\bibfnamefont{T.}~\bibnamefont{Rasing}},
  \bibinfo{journal}{Phy.\ Rev.\ Lett.} \textbf{\bibinfo{volume}{103}},
  \bibinfo{pages}{117201} (\bibinfo{year}{2009}).

\bibitem[{\citenamefont{Steil et~al.}(2011)\citenamefont{Steil, Alebrand,
  Hassdenteufel, Cinchetti, and Aeschlimann}}]{Steil2011PRB84}
\bibinfo{author}{\bibfnamefont{D.}~\bibnamefont{Steil}},
  \bibinfo{author}{\bibfnamefont{S.}~\bibnamefont{Alebrand}},
  \bibinfo{author}{\bibfnamefont{A.}~\bibnamefont{Hassdenteufel}},
  \bibinfo{author}{\bibfnamefont{M.}~\bibnamefont{Cinchetti}},
  \bibnamefont{and}
  \bibinfo{author}{\bibfnamefont{M.}~\bibnamefont{Aeschlimann}},
  \bibinfo{journal}{Phys.\ Rev.\ B} \textbf{\bibinfo{volume}{84}},
  \bibinfo{pages}{224498} (\bibinfo{year}{2011}).

\bibitem[{\citenamefont{Alebrand et~al.}(2012)\citenamefont{Alebrand,
  Hassdenteufel, Steil, Cinchetti, and Aeschlimann}}]{Alebrand2012PRB85}
\bibinfo{author}{\bibfnamefont{S.}~\bibnamefont{Alebrand}},
  \bibinfo{author}{\bibfnamefont{A.}~\bibnamefont{Hassdenteufel}},
  \bibinfo{author}{\bibfnamefont{D.}~\bibnamefont{Steil}},
  \bibinfo{author}{\bibfnamefont{M.}~\bibnamefont{Cinchetti}},
  \bibnamefont{and}
  \bibinfo{author}{\bibfnamefont{M.}~\bibnamefont{Aeschlimann}},
  \bibinfo{journal}{Phys.\ Rev.\ B} \textbf{\bibinfo{volume}{85}},
  \bibinfo{pages}{092401} (\bibinfo{year}{2012}).

\bibitem[{\citenamefont{Ostler et~al.}(2012)\citenamefont{Ostler, Barker,
  Evans, Chantrell, Atxitia, Chubykalo-Fesenko, Moussaoui, Guyader, Mengotti,
  Heyderman et~al.}}]{Ostler2012Ncomm3}
\bibinfo{author}{\bibfnamefont{T.~A.} \bibnamefont{Ostler}},
  \bibinfo{author}{\bibfnamefont{J.}~\bibnamefont{Barker}},
  \bibinfo{author}{\bibfnamefont{R.~F.~L.} \bibnamefont{Evans}},
  \bibinfo{author}{\bibfnamefont{R.~W.} \bibnamefont{Chantrell}},
  \bibinfo{author}{\bibfnamefont{U.}~\bibnamefont{Atxitia}},
  \bibinfo{author}{\bibfnamefont{O.}~\bibnamefont{Chubykalo-Fesenko}},
  \bibinfo{author}{\bibfnamefont{S.~E.} \bibnamefont{Moussaoui}},
  \bibinfo{author}{\bibfnamefont{L.~L.} \bibnamefont{Guyader}},
  \bibinfo{author}{\bibfnamefont{E.}~\bibnamefont{Mengotti}},
  \bibinfo{author}{\bibfnamefont{L.~J.} \bibnamefont{Heyderman}},
  \bibnamefont{et~al.}, \bibinfo{journal}{Nature\ Commun.}
  \textbf{\bibinfo{volume}{3}}, \bibinfo{pages}{1} (\bibinfo{year}{2012}).

\bibitem[{\citenamefont{de~la Fuente et~al.}(2004)\citenamefont{de~la Fuente,
  Arnaudas, Benito, Ciria, del Moral, Dufour, and
  Dumesnil}}]{delaFuente04JPCM16}
\bibinfo{author}{\bibfnamefont{C.}~\bibnamefont{de~la Fuente}},
  \bibinfo{author}{\bibfnamefont{J.~I.} \bibnamefont{Arnaudas}},
  \bibinfo{author}{\bibfnamefont{L.}~\bibnamefont{Benito}},
  \bibinfo{author}{\bibfnamefont{M.}~\bibnamefont{Ciria}},
  \bibinfo{author}{\bibfnamefont{A.}~\bibnamefont{del Moral}},
  \bibinfo{author}{\bibfnamefont{C.}~\bibnamefont{Dufour}}, \bibnamefont{and}
  \bibinfo{author}{\bibfnamefont{K.}~\bibnamefont{Dumesnil}},
  \bibinfo{journal}{J.\ Phys.\ Cond.\ Mat.} \textbf{\bibinfo{volume}{16}},
  \bibinfo{pages}{2959} (\bibinfo{year}{2004}).

\bibitem[{\citenamefont{Mougin et~al.}(2000)\citenamefont{Mougin, Dufour,
  Dumesnil, and Mangin}}]{Mougin00PRB62}
\bibinfo{author}{\bibfnamefont{A.}~\bibnamefont{Mougin}},
  \bibinfo{author}{\bibfnamefont{C.}~\bibnamefont{Dufour}},
  \bibinfo{author}{\bibfnamefont{K.}~\bibnamefont{Dumesnil}}, \bibnamefont{and}
  \bibinfo{author}{\bibfnamefont{P.}~\bibnamefont{Mangin}},
  \bibinfo{journal}{Phys.\ Rev.\ B} \textbf{\bibinfo{volume}{62}},
  \bibinfo{pages}{9517} (\bibinfo{year}{2000}).

\bibitem{FreeEnergyParameters} Terfenol-D is characterized by linear elastic tensor
with $c_{11}$=141~GPa, $c_{12}$=64.8~GPa and $c_{44}$=21~GPa
(Ref.~\cite{ClarkHandbookPhysChem}) and magneto-crystalline
anisotropy coefficients $K_1=-0.87$~MPa and $K_2=2.35$~MPa
(Ref.~\cite{delaFuente04JPCM16}).

\bibitem[{\citenamefont{Fashami et~al.}(2011)\citenamefont{Fashami, Roy,
  Atulasimha, and Bandyopadhyay}}]{Fashami11N22}
\bibinfo{author}{\bibfnamefont{M.~S.} \bibnamefont{Fashami}},
  \bibinfo{author}{\bibfnamefont{K.}~\bibnamefont{Roy}},
  \bibinfo{author}{\bibfnamefont{J.}~\bibnamefont{Atulasimha}},
  \bibnamefont{and}
  \bibinfo{author}{\bibfnamefont{S.}~\bibnamefont{Bandyopadhyay}},
  \bibinfo{journal}{Nanotechnology} \textbf{\bibinfo{volume}{22}},
  \bibinfo{pages}{155201} (\bibinfo{year}{2011}).

\bibitem{AcPulseDuration} In a real situation the turn-on time for the strain is
limited by the acoustic travel time across the sample. Given the
case that the thickness of Terfenol-D layer can be sufficiently
thin (Ref.~\cite{Mougin00PRB62}), the acoustic travel time falls
within the sub-picosecond time range becoming shorter than the
duration of investigated acoustic pulses.

\bibitem[{\citenamefont{Temnov}(2012)}]{Temnov12NPhoton6}
\bibinfo{author}{\bibfnamefont{V.~V.} \bibnamefont{Temnov}},
  \bibinfo{journal}{Nature\ Photon.} \textbf{\bibinfo{volume}{6}},
  \bibinfo{pages}{728} (\bibinfo{year}{2012}).

\bibitem[{\citenamefont{Temnov et~al.}(2012)\citenamefont{Temnov, Klieber,
  Nelson, Thomay, Knittel, Leitenstorfer, Makarov, Albrecht, and
  Bratschitsch}}]{TemnovAcoustoPlasmonics}
\bibinfo{author}{\bibfnamefont{V.~V.} \bibnamefont{Temnov}},
  \bibinfo{author}{\bibfnamefont{C.}~\bibnamefont{Klieber}},
  \bibinfo{author}{\bibfnamefont{K.~A.} \bibnamefont{Nelson}},
  \bibinfo{author}{\bibfnamefont{T.}~\bibnamefont{Thomay}},
  \bibinfo{author}{\bibfnamefont{V.}~\bibnamefont{Knittel}},
  \bibinfo{author}{\bibfnamefont{A.}~\bibnamefont{Leitenstorfer}},
  \bibinfo{author}{\bibfnamefont{D.}~\bibnamefont{Makarov}},
  \bibinfo{author}{\bibfnamefont{M.}~\bibnamefont{Albrecht}}, \bibnamefont{and}
  \bibinfo{author}{\bibfnamefont{R.}~\bibnamefont{Bratschitsch}},
  \bibinfo{journal}{arxiv.org:1207.6757}  (\bibinfo{year}{2012}).

\bibitem[{\citenamefont{Dehoux et~al.}(2006)\citenamefont{Dehoux, Perton,
  Chigarev, Rossignol, Rampnoux, and Audoin}}]{Dehoux06JAP100}
\bibinfo{author}{\bibfnamefont{T.}~\bibnamefont{Dehoux}},
  \bibinfo{author}{\bibfnamefont{M.}~\bibnamefont{Perton}},
  \bibinfo{author}{\bibfnamefont{N.}~\bibnamefont{Chigarev}},
  \bibinfo{author}{\bibfnamefont{C.}~\bibnamefont{Rossignol}},
  \bibinfo{author}{\bibfnamefont{J.-M.} \bibnamefont{Rampnoux}},
  \bibnamefont{and} \bibinfo{author}{\bibfnamefont{B.}~\bibnamefont{Audoin}},
  \bibinfo{journal}{J.\ Appl.\ Phys.} \textbf{\bibinfo{volume}{100}},
  \bibinfo{pages}{064318} (\bibinfo{year}{2006}).

\bibitem[{\citenamefont{Yoshimno et~al.}(1988)\citenamefont{Yoshimno, Masuda,
  Takahashi, Tsunashima, and Uchiyama}}]{Yoshino88JAP64}
\bibinfo{author}{\bibfnamefont{S.}~\bibnamefont{Yoshimno}},
  \bibinfo{author}{\bibfnamefont{M.}~\bibnamefont{Masuda}},
  \bibinfo{author}{\bibfnamefont{H.}~\bibnamefont{Takahashi}},
  \bibinfo{author}{\bibfnamefont{S.}~\bibnamefont{Tsunashima}},
  \bibnamefont{and} \bibinfo{author}{\bibfnamefont{S.}~\bibnamefont{Uchiyama}},
  \bibinfo{journal}{J.\ Appl.\ Phys.} \textbf{\bibinfo{volume}{64}},
  \bibinfo{pages}{5498} (\bibinfo{year}{1988}).

\bibitem[{\citenamefont{Dobbs}(1970)}]{Dobbs69JPCS31}
\bibinfo{author}{\bibfnamefont{E.~R.} \bibnamefont{Dobbs}},
  \bibinfo{journal}{J.\ Phys.\ Chem.\ Solids} \textbf{\bibinfo{volume}{31}},
  \bibinfo{pages}{1657} (\bibinfo{year}{1970}).

\bibitem[{\citenamefont{Thomas et~al.}(1968)\citenamefont{Thomas, Turner, and
  Bohm}}]{Thomas68PRL20}
\bibinfo{author}{\bibfnamefont{R.~L.} \bibnamefont{Thomas}},
  \bibinfo{author}{\bibfnamefont{G.}~\bibnamefont{Turner}}, \bibnamefont{and}
  \bibinfo{author}{\bibfnamefont{H.~V.} \bibnamefont{Bohm}},
  \bibinfo{journal}{Phys.\ Rev.\ Lett.} \textbf{\bibinfo{volume}{20}},
  \bibinfo{pages}{207} (\bibinfo{year}{1968}).

\end{thebibliography}



\end{document}